\documentclass[prl,twocolumn,showpacs,preprintnumbers,amsmath,amssymb]{revtex4}
\usepackage{txfonts}
\usepackage{epsfig,amsmath,amssymb} 
\usepackage[all]{xy}
\usepackage{ifpdf} 
\usepackage{graphicx}   
\usepackage{amsfonts}
\usepackage{makeidx}  
\usepackage{epigraph}
\usepackage{etoolbox}
\makeatletter
\patchcmd{\epigraph}{\@epitext{#1}}{\itshape\@epitext{#1}}{}{}
\makeatother

\newcommand{\qed}{\hfill \mbox{\raggedright \rule{.07in}{.1in}}}

\newcommand{\ket}[1]{\left | #1 \right\rangle}

\newcommand{\qq}[1]{``#1"}

\begin{document}

\title{Why we need to quantise everything, including gravity} 

\author{C. Marletto$^{a}$ and V. Vedral $^{a,b}$
	\\ {\small $^{a}$ Physics Department, University of Oxford} 
	\\{\small $^{b}$Centre for Quantum Technologies, National University of Singapore}}

\date{February 2016}

\begin{abstract}
	
\noindent There is a long-standing debate about whether gravity should be quantised. A powerful line of argument in favour of quantum gravity considers models of hybrid systems consisting of coupled quantum-classical sectors. The conclusion is that such models are inconsistent: either the quantum sector's defining properties necessarily spread to the classical sector, or they are violated. These arguments have a long history, starting with the debates about the quantum nature of the electromagnetic fields in the early days of quantum theory. Yet, they have limited scope because they rely on particular dynamical models obeying restrictive conditions, such as unitarity. In this paper we propose a radically new, more general argument, relying on less restrictive assumptions. The key feature is an information-theoretic characterisation of both sectors, including their interaction, via constraints on copying operations. These operations are necessary for the existence of observables in any physical theory, because they constitute the most general representation of measurement interactions. Remarkably, our argument is formulated without resorting to particular dynamical models, thus being applicable to any hybrid system, even those ruled by \qq{post-quantum} theories. Its conclusion is also compatible with partially quantum systems, such as those that exhibit features like complementarity, but may lack others, such as entanglement. As an example, we consider a hybrid system of qubits and rebits. Surprisingly, despite the rebit's lack of complex amplitudes, the signature quantum protocols such as teleportation are still possible.

\bigskip

\end{abstract}

\maketitle

\noindent Quantum theory and general relativity are the two most fundamental theories of physics;  yet we know that one of them, or both, might ultimately have to be modified into a more general theory, which will resolve their clash. 

\noindent A possible route to that general theory is the quantisation of gravity, in the same spirit as other field theories; yet, whether or not this quantisation is necessary is still controversial. A number of arguments have been proposed in favour of it, \cite{DeW}, \cite{FEY}, \cite{TER}, to show that, as Feynman put it, \qq{We are in trouble if we believe in quantum mechanics, but we do not quantize gravitational theory.} \cite{FEY}. 

These arguments all address the more general problem of whether a hybrid system composed of a quantum and a classical sectors, coupled via some interaction, is possible -- which goes well beyond the domain of quantum gravity. It was already considered by Heisenberg in the context of the coupling between electromagnetic fields and quantum charges, \cite{HEI}, and it has been an open question ever since \cite{BOHRING}. In such arguments, the quantum sector is required to obey unitary quantum theory. The classical sector instead is required to satisfy a number of different conditions, capturing some notion of classicality: for instance, in  \cite{DeW} it is required to have observables which can all be sharp simultaneously; in \cite{TER} to obey a Liouvillian dynamics in phase space; in \cite{TER1} to be described by a Wigner function whose evolution is generated by Moyal brackets. In all such cases, the conclusion is that the model describing the hybrid classical-quantum system is inconsistent. In particular, either the quantum sector must spread its features to the classical sector; or the uncertainty relations are violated on the quantum sector.

\noindent Now, these arguments have two problems. First, they all rely on adopting a specific hybrid dynamical formalism, capturing both quantum and classical dynamics, thus having limited scope and applicability. Also, some of them assume particular couplings between the classical and the quantum sectors or particular measuring procedures, following Heisenberg's logic  \cite{PERES, RUS}. 

\noindent In addition, the dynamics on the composite system is frequently assumed to be unitary, i.e., linear,  deterministic and reversible. This is too strong an assumption, since it assumes the fundamental trait of unitary quantum dynamics on the composite system, thus almost forcing that trait to hold on the classical sector. 

In this paper we propose a radically new argument which is free of these shortcomings, formulated exclusively in information-theoretic terms. It is independent of specific details of the dynamical theories or measurement procedures that couple the two sectors; and the quantum sector is modelled in such a way that it includes even systems that obey post-quantum theories. The information-theoretic nature of the argument allows it to be applicable to more general cases, where the classical sector need not be the gravitational field, but a generic non-quantum system, e.g. a simple harmonic oscillator or a classical bit. We shall nonetheless refer to the two sectors as \qq{quantum} and \qq{classical}, with the understanding that those notions are generalised as detailed in our model. 
 
In this more general form, the argument has new interesting consequences. In particular, it is compatible with the classical sector not obeying the full quantum theory. As an example, we consider a hybrid system, made of qubits and rebits, \cite{WOOT}. A rebit has complementary observables, but its dynamics is restricted to states that have only real amplitudes. Yet, in such hybrid systems superdense coding and teleportation are still possible. This provides a deeper insight into what \qq{quantising} a system really means; it also highlights an interesting difference between ways of implementing the swap operation, which is not manifest when the two systems are of the same kind. \\

To formulate the argument independently of dynamics, we adopt the approach of the constructor theory of information, \cite{DEMA, MA}. The advantage is that there is no need to assume any specific underlying dynamical model and the assumptions can be cast in general information-theoretic terms. 

\noindent In constructor theory physical systems are {\sl substrates} –- i.e., physical systems on which physical transformations, or {\sl tasks}, can be performed. An {\sl attribute} {\bf n} is the set of all states where the substrate has a given property. For instance, the set of all quantum states of a qubit where a given projector is sharp with value $1$ is an attribute. A task is the specification of a physical transformation, in terms of input/output pairs of attributes. For example, the NOT task on the attributes ${\bf 0 }$, ${\bf 1}$ is written as $\{ {\bf 0}\rightarrow{\bf 1}\;,\;\;{\bf 1}\rightarrow{\bf 0}\}$.

\noindent A task is {\sl impossible} if the laws of physics impose a limit to how accurately it can be performed. Unitary quantum theory, for instance, requires the task of cloning sets of non-orthogonal quantum states to be impossible \cite{NOCLO}. Otherwise, the task is {\sl possible}: there can be arbitrarily good approximations to a {\sl constructor} for it, which is defined as a substrate that, whenever presented with the substrates in any of the input attributes of the task, delivers them in one of the corresponding output attributes, and, crucially, retains the property of doing it again. In quantum information, gates for computational tasks are example of constructors \cite{DEUGATE}.

In constructor theory one can provide a powerful information-theoretic characterisation of the classical and quantum sectors, without specifying particular dynamical laws, by regarding them as different classes of substrates, defined as follows \cite{DEMA}.
First, one defines an \qq{information medium} as a substrate with a set of attributes $X$, called {\sl information variable}, with the property that the following tasks are possible: 

\begin{equation}
\bigcup_{{x}\in X}\left\{({\bf x},{\bf x_0})\rightarrow ({\bf x},{\bf x}) \right\}\;, \label{kk}
\end{equation}

\begin{equation}
\bigcup_{{x}\in X}\left\{{\bf x}\rightarrow \Pi({\bf x}) \right\} \label{dd}
\end{equation}

\noindent for all permutation $\Pi$ on the set of labels of the attributes in $X$ and some blank attribute ${\bf x_0}\in X$.

\noindent The former task corresponds to \qq{copying}, or cloning, the attributes of the first replica of the substrate onto the second, target, substrate; the latter, for a particular $\Pi$, corresponds to a logically reversible computation (which need not require it to be realised in a physically reversible way). So, an information medium is a substrate that can be used for classical information processing (but could, in general, be used for more). For example, a qubit is an information medium with any set of two orthogonal quantum states.  

\noindent Any two information media (e.g. a photon and an electron) must satisfy an \qq{interoperability principle} \cite{DEMA}, which expresses elegantly the intuitive property that classical information must be copiable from one information medium to any other, irrespective of their physical details. Specifically, if $S_1$ and $S_2$ are information media, respectively with information variable $X_1$ and $X_2$, their composite system $S_1 \oplus S_2$ is an information medium with information variable $X_1\times X_2$, where $\times$ denotes the Cartesian product of sets. This requires the task of copying information variables (as in eq. \eqref{kk}) from one to the other to be possible.
 
\noindent With these tools one can express information-theoretic concepts such as measuring and distinguishing, without resorting to formal concepts such as orthogonality, linearity or unitarity. This is the key feature that will allow our argument to be independent of particular dynamical models. The variable $X$ is {\sl distinguishable} if the task

\begin{equation}
\bigcup_{{x}\in X}\left\{{\bf x}\rightarrow {\bf q_x} \right\} \nonumber
\end{equation}

\noindent is possible, where the variable $\{{\bf q_x}\}$ is some information variable. If the variable $\{{\bf x_0}, {\bf x_1}\}$ is distinguishable, we say that the attribute ${\bf x_0}$ is distinguishable from ${\bf x_1}$, denoted as ${\bf x_0}\perp{\bf x_1}$.  This notion of distinguishability allows one to generalise the orthogonal complement of a vector space: for any attribute ${\bf n}$ define the attribute $\bar {\bf n}$ as the union of all attributes that are distinguishable from ${\bf n}$. 

\noindent An {\sl observable} is an information variable whose attributes ${\bf x}$ have the property that $\bar{\bar{\bf x}}={\bf x}$; this notion generalises that of a quantum observable. An observable $X$ is said to be {\sl sharp} on a substrate, with value $x$, if the substrates is in a state $\xi$ that belongs to one of the attributes ${\bf x}\in X$. 
A special case of the distinguishing task is the perfect measurement task: 

\begin{equation}
\bigcup_{{x}\in X}\left\{({\bf x}, {\bf x_0})\rightarrow ({\bf x}, {\bf p_x}) \right\} \label{PD}
\end{equation} where the first substrate is the \qq{source} and the second the \qq{target}. From the interoperability principle, it follows that the above task must be possible for any information variable. A {\sl measurer of $X$} is any constructor that can perform the above task, for some choice of the {\sl output variable} $\{{\bf p}_x\}$.
In particular, \cite{DEMA}, a measurer of an observable $X$ has the property that if the output variable is sharp with value ${\bf p}_x$ on the target, then the observable $X$ must be sharp on the source in input, with value $x$. 

We can now formulate the argument for quantisation using these information-theoretic tools only. One starts by characterising the two sectors, as follows.  

The classical sector $S_C$ is modelled as an information medium with the property that the union of all of its information observables is an information observable $T$. This means that its observables can all be simultaneously sharp. 

\noindent To model the quantum sector $S_Q$ we resort to the constructor-theoretic notion of a {\sl superinformation medium} --  an information medium with at least two {\sl disjoint} information observables $X$ and $Z$, with the property that $X\cup Z$ is {\sl not} an information variable. A qubit, for example, is a superinformation medium with observables $X=\{{\bf x_1},{\bf x_2}\}$ and $Z=\{{\bf z_1},{\bf z_2}\}$, where $z_i$ labels the attribute of being in the i-th eigenstate of the $Z$ component of the qubit, and likewise for $X$.  This is because the variable $\{{\bf z_1},{\bf z_2}, {\bf x_1}, {\bf x_2}\}$ is not copiable perfectly under unitary quantum theory, and therefore it is not an information variable. On the other hand, a system obeying e.g. Spekkens' toy model \cite{SPEK} is not a superinformation medium, because the reason why attributes in different observables cannot be cloned is that their intersection is non-empty in the space of states (in that they correspond to an uncertainty in preparation). 

\noindent In addition, we require that $S_Q$ has the property that a perfect measurer of $Z$ is also capable of distinguishing $X$, as in \eqref{dd}, and vice versa. This is true in quantum theory, where a CNOT in the $X$ basis can be used perfectly to discriminate the states in the $Z$ basis.

We shall now proceed to demonstrate the main result of our paper -- namely that the classical sector $S_C$ must itself be a superinformation medium with at least two observables which cannot be sharp simultaneously. Informally speaking, the generalised \qq{complementarity} feature of the quantum sector must spread to the classical sector.  

\noindent For simplicity, we assume that the all the information observables are binary: $T=\{{\bf t_1}, {\bf t_2}\}$, $Z=\{{\bf z_1}, {\bf z_2}\}$, $X=\{{\bf x_1}, {\bf x_2}\}$. As shown in \cite{DEMA}, the attributes ${\bf x}$ in $X$ generalise quantum superpositions or mixtures of the eigenstates of $Z$. In particular, define the attribute $u_Z\doteq \bigcup_{z\in Z}{\bf z}$; one then has that $\forall {x}, z$: ${\bf x}\not \perp {\bf z}$, ${\bf x}\cap {\bf z}=\{\emptyset\}$, and ${\bf x}\subset \bar{\bar{u}}_Z$ -- which corresponds to the fact that each ${\bf x}$ generalises the notion of a quantum state that is in the \qq{span} of the eigenstates of $Z$, but is not an eigenstate of $Z$.

\noindent A particular realisation of a measurer of $Z$ is a constructor for the following task on the joint substrate $S_Q\oplus S_C$: $$\{({\bf x_1},{\bf t_1})\rightarrow({\bf x_1},{\bf t_1}), ({\bf x_2},{\bf t_1})\rightarrow({\bf x_2},{\bf t_2}) \}\;.$$ This measurer must exist as a consequence of interoperability of information media. 

\noindent Now, suppose this measurer of $Z$ is presented with ${\bf x}$ where $X$ is sharp. We have: 
$$\{({\bf x_1},{\bf t_1})\rightarrow {\bf p_{+}}, ({\bf x_2},{\bf t_2})\rightarrow {\bf p_{-}}\}\;,$$ where, by the requirement that the measurer of $Z$ can also distinguish $X$, $\{{\bf p_{+}}, {\bf p_{-}}\}$ is an information variable. 

\noindent Let us define an observable $S$ on the composite system, which once again must exist by the interoperability of information, whose physical interpretation is \qq{whether the first substrate has the same label $z$ as the second substrate}.  $S$ generalises what in quantum theory would be the projector for being in the symmetric subspace. By relabelling $t_1$ as $z_1$ and $t_2$ as $z_2$, it follows that $S$ must be sharp in both ${\bf p_{+}}$ and ${\bf p_{-}}$, with value \qq{Yes}, \cite{MA}.

\noindent  A particular way of measuring $S$ is to apply a measurer of $T$ to $S_C$, with target $S_Q$, defined so that: 

\begin{eqnarray}
({\bf z_1},{\bf t_1})\rightarrow({\bf z_1},{\bf t_1})\nonumber \\
({\bf z_2},{\bf t_2})\rightarrow({\bf z_1},{\bf t_2})\nonumber \\
({\bf z_2},{\bf t_1})\rightarrow({\bf z_2},{\bf t_1})\nonumber \\
({\bf z_1},{\bf t_2})\rightarrow({\bf z_2},{\bf t_2})\nonumber \\
\end{eqnarray}

\noindent where $Z$ being sharp on the target $S_Q$ with value $z_1$ means \qq{yes} and with value $z_2$ means \qq{no}. Therefore, a measurer of $T$ applied to $S_C\oplus S_Q$ prepared with the attributes ${\bf p_+}$ or ${\bf p_-}$ delivers in output $S_Q$ with $Z$ sharp with value $z_1$:

\begin{eqnarray}
{\bf p_+}\rightarrow({\bf z_1},{\bf r_1})\nonumber \\
{\bf p_-}\rightarrow({\bf z_1},{\bf r_2}) \nonumber
\end{eqnarray}

\noindent where ${\bf r_1}$ and ${\bf r_2}$ are some attributes of the classical sector $S_C$ -- which whose existence is a consequence of this argument. 

\begin{figure}[h]
	\centering
	\includegraphics[scale=0.4]{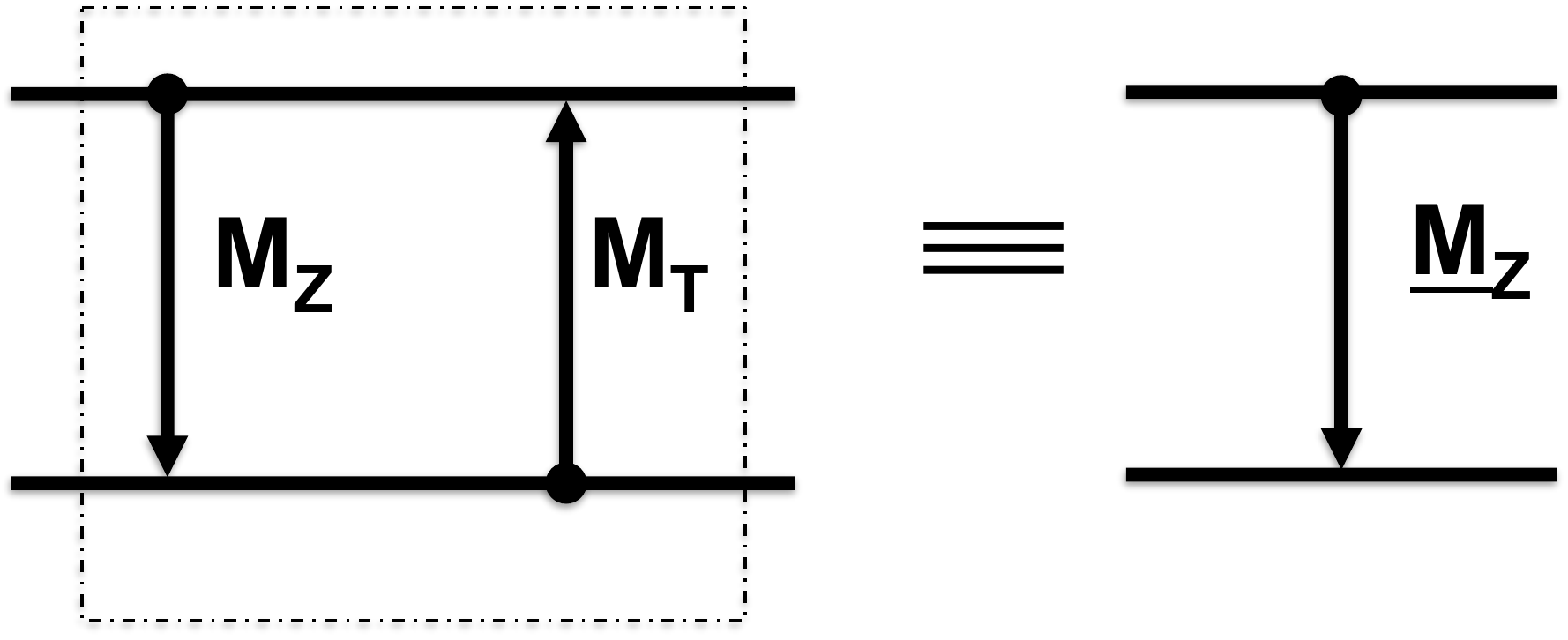} 
	\caption{Equivalence of measurers. A measurer $M_Z$ of $Z$ acting on $S_Q$ composed with a measurer $M_T$ of $T$ on $S_C$ is equivalent to a different realisation $\underbar M_Z$of a measurer of $Z$ on $ S_Q$.}
	
\end{figure}

\noindent Now, crucially, the two attributes ${\bf r_1}$ and ${\bf r_2}$ can be shown to constitute an information observable, which cannot be simultaneously sharp when $T$ is sharp. Specifically: 

The observable $T$ cannot be sharp in any of the states in either ${\bf r_1}$ or ${\bf r_2}$, \cite{MA}: $\forall {\bf t} \in T$, ${\bf r_1}\cap {\bf t}$ and ${\bf r_2}\cap {\bf t}$ are empty. For suppose, say, ${\bf r_1}\cap {\bf t_1}$ included some state $s$. Then, by definition of measurer of $Z$, $Z$ should be sharp in input on $S_Q$, with value $z_1$; but this contradicts the defining property of superinformation media, that ${\bf z_1}$ and ${\bf x_1}$ have empty intersection.  Likewise for the other attributes. In addition, ${\bf r_1} \not \perp {\bf t}$ or ${\bf r_1} \not \perp {\bf t}$, because otherwise, again, ${\bf x_1}$ and ${\bf x_2}$ should be distinguishable from some of the ${\bf z}$'s, contrary to assumption.  Also, upon defining $u_T=\bigcup_{t\in T}{\bf t}$, both ${\bf r_1}$ and ${\bf r_2}$ are included in $\bar{\bar{u}}_T$ -- i.e., they must belong to the generalisation of what in quantum theory would be called the span of the quantum states in the $T$ basis. Note that this conclusion would follow even without assuming the additional property that a measurer of $Z$ can also distinguish the attributes in $X$. This would correspond to a weaker result, showing that the classical sector must at least exhibit the equivalent of mixtures –- but need not have two complementary observables. 

In addition, composing a measurer of $T$ with a measurer of $Z$ as above still gives a measurer of $Z$ (see the figure 1). By requiring that any measurer of $Z$ can also distinguish $X$ (i.e. it can perform the task \eqref{dd} for $X$), one has that the ${{\bf r_1}, {\bf r_2}}$ must be an information observable, too.  

\noindent Hence $S_C$ must have two \qq{complementary observables}: $T=\{{\bf t_1},{\bf t_2}\}$ and $R=\{{\bf r_1},{\bf r_2}\}$, thus being a superinformation medium with those two observables. This concludes our argument that $S_C$ must also be \qq{quantised}. 

Note that it is impossible for ${\bf r_1}$ and ${\bf r_2}$ to be ‘statistical mixtures’ of ${\bf t_1}$ and ${\bf t_2}$, because they are perfectly distinguishable from one another. In addition, our argument rules out the case where ${\bf r_1}$ and ${\bf r_2}$ each correspond to sets of states intersecting with ${\bf t_1}$ and ${\bf t_2}$, e.g. because they represent some irreducible limitation to resolution in the space of states, as in \cite{SPEK}.   

\noindent Remarkably, this argument does not assume unitarity or reversibility of the dynamical laws, which was assumed in previous arguments. One only uses the definition of an observable, of a measurer, and the fact that the measurer acts deterministically on all states (any state in input produces the same state in output every time the measurer acts upon it). Hence, the argument does not apply to stochastic, collapse-based variants of quantum theory, e.g. \cite{GRW, WIN}. Note also that a key assumption is that a measurer of $Z$ can also distinguish $X$: therefore, the argument would not apply to theories claiming that quantum gravity has no observable effects, e.g. \cite{NOB1, NOB}.

\noindent The same result can be achieved by supposing, as in \cite{FEY}, that the measurement interaction is physically reversible on the composite substrate $S_C\oplus S_Q$, when $S_Q$ is prepared in any of the attributes in the joint variable $X \cup Z$. This is a stronger assumption, which implies our assumption that the measurer of $X$ is capable of distinguishing $Z$. Reversibility in this sense holds in quantum information theory, but need not be true in general.

Our argument does not require $S_C$ to obey strictly quantum theory, because our assumptions allow for more general cases. However, we do believe that more specific features of bipartite quantum systems, such as the maximum violation of the Bell inequality, can also be derived by adding further information-theoretic constraints \cite{MA}, in the same spirit as \cite{PAW}.

As an example of a more general hybrid system, we shall consider systems consisting of two different superinformation media: qubits and rebits.

\noindent A rebit is a superinformation medium which is obtained from a qubit by requiring that the information observables (as defined above) are only the physical quantities represented by real-valued operators. We interpret this as meaning that all states in the Block sphere can be {\sl prepared}, but only a subset of pairs of orthogonal pure states can be discriminated; precisely those that can be represented as 2-dimensional vectors on the real Hilbert space,  $\ket{\psi}=a\ket{0}+b\ket{1}$ where $a$ and $b$ are real amplitudes whose squares add up to $1$, and $\ket{0}$ and $\ket{1}$ are the eigenvectors of the observable $Z$. Complex amplitudes, in other words, are not observable. A single rebit can be used to perform interference experiments, because it has two non-commuting observables: $X$ and $Z$. But what about other hallmark quantum-information processing tasks?

\noindent Superdense coding \cite{BEN} can be implemented with just two rebits, given that it involves only measuring $X, Z$ and $iY$, which are real observables.  Remarkably, a hybrid system of qubits and rebits can also be used to perform teleportation \cite{JOS}. Specifically, one can use a qubit, initially prepared in the state ${\ket \psi}$ to be teleported; two rebits, prepared in a Bell state, and another ancillary qubit, prepared in some blank state $\ket{0}$. The crucial step in the teleportation protocol is to teleport the state to the second rebit, by using a Bell measurement on the composite system of the first qubit and the other rebit. When the state $\ket{\psi}$ is teleported to the second rebit, its phases are inaccessible, as only real-valued operators can be measured. However, one can swap the state on the rebit with that of the other, ancillary, qubit. This is possible because the swap can be accomplished as a sequence of three CNOT gates in the Z basis operating between the rebit and the qubit. These are precisely the copying operations (i.e. measurements) that we assume to be allowed on the two subsystems. The readout of the teleported state is then performed by measurements on the qubit. 

These hybrid systems of qubits and rebits bring out an interesting asymmetry between superdense coding and teleportation. They also display a crucial difference between swapping physically two systems (e.g. moving the two carriers of information in space) and performing the logical swap of their states (which can be realised by a sequence of three CNOT gates, operating on the quantum degrees of freedom of the carriers). These two operations are not the same for a system made of a qubit and a rebit; but they are the same for a system of two qubits. Since the logical swap might require different resources to the physical swap, in a qubit-rebit system the usual assumption about the swap being a trivial operation might not apply, thus leading to a different notion of what the elementary/complex operations are.

Our constructor-information-theoretic proposal radically changes the approach to arguments for quantisation, emancipating them from specific, narrow and overly-constrained dynamical models. Our assumptions are far less restrictive, in that they do not rely on unitarity or reversibility. They are also cast in an elegant, exact information-theoretic form –- exclusively in terms of the possibility/impossibility of certain copying tasks on the two sectors, which provides a more general notion of being \qq{quantum}. This is achieved via the constructor theory of information, which provides a powerful new approach to exploring the interaction of systems obeying different physical theories, even post-quantum. Our argument thus provides  radically new foundations, tools and content for the debate about quantisation of a classical system interacting with a quantum system. These are applicable not only in the context of quantum gravity, but also in other fields where the quantum-classical boundaries need to be explored: from quantum information to electrodynamics; from cosmology to quantum thermodynamics.

\textit{Acknowledgments}: 
CM's research was supported by the Templeton World Charity Foundation. VV thanks the Oxford Martin School, Wolfson College and the University of Oxford, the Leverhulme Trust (UK), the John Templeton Foundation, the EU Collaborative Project TherMiQ (Grant Agreement 618074), the COST Action MP1209, the EPSRC (UK) and the Ministry of Manpower (Singapore). This research is also supported by the National Research Foundation, Prime Minister’s Office, Singapore, under its Competitive Research Programme (CRP Award No. NRF- CRP14-2014-02) and administered by Centre for Quantum Technologies, National University of Singapore.

\end{document}